# Understanding the detectability of potential changes to the 100-year peak storm surge

**Robert L. Ceres[1], Chris E. Forest[1-3], and Klaus Keller[2-4]**

[1]Department of Meteorology and Atmospheric Science, The Pennsylvania State University, University Park, Pennsylvania, USA.

[2]Department of Geosciences, The Pennsylvania State University, University Park, Pennsylvania 16803, USA.

[3]Earth and Environmental Science Institute, The Pennsylvania State University, University Park, Pennsylvania 16803, USA.

[4]Department of Engineering and Public Policy, Carnegie Mellon University, Pittsburgh, Pennsylvania 15213, USA.

Corresponding author: Robert Ceres (RobCeres@psu.edu)

**Abstract**

In many coastal communities, the risks driven by storm surges are motivating substantial investments in flood risk management. The design of adaptive risk management strategies, however, hinges on the ability to detect future changes in storm surge statistics. Previous studies have used observations to identify changes in past storm surge statistics. Here, we focus on the simple and decision-relevant question: How fast can we learn from past and potential future storm surge observations about changes in future statistics? Using Observing System Simulation Experiments, we quantify the time required to detect changes in the probability of extreme storm surge events. We estimate low probabilities of detection when substantial but gradual changes to the 100-year storm surge occur. As a result, policy makers may underestimate considerable increases in storm surge risk over the typically long lifespans of major infrastructure projects.

# 1 Introduction

Hurricanes Katrina and Sandy recently caused thousands of deaths and hundreds of billions in property damages (Kunz et al. 2013). These storms highlight the difficulties in perceiving, estimating, and preparing for the potential danger of storm surges. Within major metropolitan regions, developers and planners are proposing, designing, and constructing costly and long-lived infrastructure to address storm surge risks (e.g. NYC Economic Development Corporation 2014). An improved understanding of future storm surge risks can inform decisions about risk management strategies. Previous studies using observations to identify changes in the frequency or magnitude of extreme storm surges and other climate-driven weather phenomena have noted the difficulties in detecting these changes (e.g., Menéndez and Woodworth 2010; Grinsted et al. 2012, 2013; Reich et al. 2014; Talke et al. 2014a; Stephenson et al. 2015; Huang et al. 2016; Vousdoukas et al. 2016). Here we explore and quantify the time required to detect changes to extreme storm surges.

In the United States, tropical and extratropical cyclone storm surges cause most extreme weather damages (Harris 1963; Blake et al. 2007), and storm surge damages have been increasing (Blake et al. 2007; Smith and Katz 2013). Cyclone storm surges result from a combination of wind-driven transport over the ocean and from rain over land. These, in turn, are related to sea level, storm frequency, intensity, size, and track (Harris 1963). The IPCC's Climate Change 2014 Synthesis Report Summary for Policymakers (Pachauri et al. 2014) states: "Global mean sea level rise will continue during the 21st century, very likely at a faster rate than observed from 1971 to 2010."

The impacts of other climate-related drivers of storm surge risk remains uncertain (Kunkel et al. 2013). Several factors complicate the detection of climate driven changes in storm surges. For one, tide gauge data may not span the long time period that is needed for detection (Kunkel et al. 2013). In addition, current climate models face challenges in realistically resolving hurricanes (Knutson et al. 2008; Bender et al. 2010; Kunkel et al. 2013; Bacmeister et al. 2014). Nevertheless, many studies suggest that the global frequency of cyclones is likely to decrease (or stay the same), and the frequency and intensity of the most intense cyclones are likely to increase (Emanuel 1987; Beniston et al. 2007; Knutson et al. 2008; Bender et al. 2010; Reed et al. 2015; Little et al. 2015; Kossin 2017). Pachauri et al. (2014) states, "In urban areas climate change is projected to increase risks for people, assets, economies and ecosystems, including risks from heat stress, storms and extreme precipitation, inland and coastal flooding, landslides, air pollution, drought, water scarcity, sea level rise and storm surges (very high confidence)". Given these expected changes, researchers are analyzing how storm surge characteristics are changing over time and depend on climate variables (e.g., Fischbach 2010; Cheng et al. 2014; Vousdoukas et al. 2016).

Independent of these climate related dependencies, researchers are separately using tide gauge observations to estimate trends in sea level and storm surge (Fischbach 2010; Hoffman et al. 2010; Maloney and Preston 2014; Kopp et al. 2015; Muis et al. 2016). Tide gauge observations provide a measurable characteristic of storms that correlates well with storm related economic losses and can often provide a long-term consistent data source (Grinsted et al. 2012; Talke et al. 2014a, b). However, in the specific case of storm surges, it is unclear how quickly changes in storm surge statistics might be detected when including future observations.

Researchers have proposed (e.g., Lempert et al. 1996, 2012; Haasnoot et al. 2013), and policy makers are embracing dynamic adaptive strategies to adjust risk management strategies as new information becomes available (e.g. NYC Economic Development Corporation 2014). The design of such adaptive strategies is complicated, for example, by the deep uncertainty surrounding potential changes in future storm surges, the time needed to plan and implement the often large required investments, and the often long lifetimes of surge protection infrastructure. The implementation of adaptive strategies often hinges on the detection of signposts to trigger a change in strategy (e.g., Lempert et al. 1996; Kwakkel et al. 2015).

In the US, the 100-year flood level (also known as the "base flood" in FEMA 2015) is widely used as an indicator to inform the assessments of flooding risk and the need for risk mitigation (U.S. CFR 725 Executive Orders 11988 1988, 2015; Bellomo et al. 1999; FEMA 2015). The 100-year storm surge is defined as the surge level (also referred to as the return level) that will, on average in a statistically stationary system, be exceeded each 100 years (the return period) with a one percent chance of being exceeded during a given year (the annual exceedance probability) (Gilleland and Katz 2011; Cooley 2013). The return level concept can be confusing in a nonstationary system (Cooley 2013). For the remainder of this article, we use 100-year storm surge to refer to the "effective return level" (as defined in Gilleland and Katz 2011). Specifically, this is the potentially time-dependent surge level expected to be exceeded with a one percent probability in a one-year period. In practice, "base flood" 100-year return levels are often expressed as a line (or threshold) with no uncertainty (e.g., U.S. CFR 725 Executive Orders 11988 1988, 2015; NYC Economic Development Corporation 2014; FEMA 2015; Kaplan et al. 2016). The Federal Emergency Management Agency periodically updates and publishes flood maps showing the base flood line based on historic data and computer models (FEMA 2015). Regions above this line may be interpreted as being safe, whereas regions below the line are considered at risk and subject to risk mitigation requirements (FEMA 2005). Furthermore, in the US, risk mitigation strategies are often intended to withstand events up to the 100-year return level without consideration for more extreme, less probable events (FEMA 2005; Kaplan et al. 2016).

We adopt this familiar 100-year storm surge metric (without uncertainty) as a useful starting point. We then ask: how long would it take to detect a change in the 100-year storm surge? This pragmatic interpretation shifts uncertainty from estimated 100-year surge levels to a probabilistic estimate of time required to detect changes in the 100-year storm surge reliably. By knowing the delay in detecting this signpost and the manner in which signpost metrics should behave over time, we can help inform the design of adaptive risk management strategies (e.g., Lempert et al. 2012; Kwakkel et al. 2015; Buchanan et al. 2016), such as those currently proposed for NYC (NYC Economic Development Corporation 2014). It is therefore important to know how long it will take to realize that risk levels have changed.

To quantify detection times, we use an observation system simulation experiment (OSSE). Specifically, we generate and analyze simulated observations from OSSE nature states with prescribed changes to statistical parameters. Based on these observations, we calculate the frequency of detection over a range of detection intervals and explore how estimated 100-year storm surges change over those intervals.

Although the peak water-level results from the combination of local sea level, celestial tide, and storm-driven surge, we focus here on the storm-driven surge. This focus is motivated by the deep uncertainty associated with estimates of future surges (Oddo et al. 2017) and because the uncertainties surrounding future surges can be the major driver of future flooding risks (Wong and Keller (Wong and Keller 2017).

**2. Data, model, and methods**

As an illustrative and consistency purposes, we consider the tide gauge readings at The Battery in lower Manhattan (NOAA 2015). We remove celestial tides and local sea level rise, and extract the annual block maximum reading from each consecutive one-year sequence of data (see supplementary materials). We next use extreme value theory (Fisher and Tippett 1928; Coles 2001) to statistically model the frequency and intensity of extreme, but rare, events. Specifically, we use a maximum likelihood estimate (Nelder and Mead 1965) for the parameters in the Generalized Extreme Value (GEV) distribution (equation 1). (See supplemental material for justification of using this approach.)

The GEV probability density function (equation 1),

$$f(x; \mu, \sigma, \xi) = \frac{1}{\sigma}\left[1 + \xi\left(\frac{x-\mu}{\sigma}\right)\right]^{\left(\frac{-1}{\xi}\right)-1} exp\left\{-\left[1 + \xi\left(\frac{x-\mu}{\sigma}\right)\right]^{\frac{-1}{\xi}}\right\}, \quad (1)$$

for return level, $f(x)$, is defined by three parameters: location $\mu$, scale $\sigma$, and shape $\xi$. The location parameter ($\mu$) corresponds to the expected value of the distribution with respect to the block maxima storm surge heights. A change $\mu$ to, $\mu + \Delta\mu$, shifts every point in distribution by $\Delta\mu$. Alternatively, the relationship between $\mu$ and $f(x)$ can be expressed as:

$$f(x; \mu + \Delta\mu, \sigma, \xi) = f(x + \Delta\mu; \mu, \sigma, \xi). \quad (2)$$

The scale parameter ($\sigma$) describes the width of the distribution. The shape parameter ($\xi$) is associated with the skewness of the distribution. Changes in future cyclone frequency, storm size, intensity, tracks, or speeds may lead to changes in any of these parameters. For example, an increase in the frequency of storms generating all magnitudes of storm surges will shift $\mu$ (independent of the effect of local sea level, which has been removed) in accordance with equation (2). Similarly, nonuniform changes to storm characteristics could result in changes to $\sigma$ or $\xi$.

We calculate the expected exceedance levels over any time period using the maximum likelihood parameter estimates. The parameters for The Battery tide gauge ($\mu_0$ = 0.936 m, $\sigma_0$ = 0.206 m, and $\xi_0$ = 0.232) produce an estimated 100-year storm surge of 2.62 m above MHHW datum, approximately the storm surge observed during hurricane Sandy (Hall and Sobel 2013). This 100-year surge (and the corresponding 95% confidence interval from 1.6 to 3.7m) is consistent with other estimates of The Battery's 100-year surge (n.b., they are not provided in the papers) (Talke et al. 2014b; Lopeman 2015). Our GEV parameters are broadly consistent with one pre-Sandy estimate ($\mu$ = 0.935 m, $\sigma$ = 0.260 m, and $\xi$ = 0.030) (Kirshen et al. 2007). Note that Sandy's surge exceeded the study's estimated 2.05 m 100-year surge, which, if included, would be expected to increase $\sigma$ or $\xi$.

GEV estimation methods often assume time-invariant properties are generating the analyzed time series (Cooley 2013). This assumption, however, may be violated (van Dantzig 1956; Kunkel et al. 2013; Grinsted et al. 2013; Little et al. 2015; Huang et al. 2016). We hence explore the performance differences between stationary and nonstationary analyses.

To impose changes in the 100-year surge, we can change $\mu$, $\sigma$, or $\xi$ (see Fig. 1). Changing $\sigma$, from 0.206 to 0.323, for example, results in a one meter increase in the projected 100-year surge, increasing the probability for all annual maxima greater than ~1.5 meters. Similarly, changing $\mu$ from 0.936 to 1.936 increases the projected 100-year surge by one meter, increasing the probability for annual storm surge maxima greater than ~1.8 meters. Visual inspection of Figure 1 might suggest that the increase in $\mu$ is much worse (in terms of higher probabilities for surges greater than the current 100-year surge) than the increase in $\sigma$. Closer inspection, however, reveals that the reverse is true for very large storm surges. That is, the change in $\sigma$ produces the higher percentage of surges greater than ~4.5 meters, including surge levels that would exceed the 19 foot protection heights currently proposed for lower Manhattan (i.e., above the NAVD88 datum, corresponding to 5.6 m above MHHW datum) (NYC Economic Development Corporation 2014).

Changes to $\xi$ imply that the behavior of the extreme tail is changing, but estimating $\xi$ accurately is nontrivial (Coles 2001). Following the common practice in nonstationary GEV analysis (Coles 2001; Kharin and Zwiers 2005), we focus on the relatively simpler detection problem with a stationary $\xi$. A more detailed discussion of $\xi$ is included in the supplement.

In this paper, we use "nature state" to mean a set of GEV parameters with prescribed changes over time, "nature run" to mean a simulation generated using a specific nature state, and "simulated observation" to mean an estimated 100-year surge calculated from a nature run. All nature states (see supplemental Table S1) use baseline parameters, ($\mu_0$, $\sigma_0$, and $\xi_0$). For each experiment, we generate 100,000 nature runs of a 200-year time series of annual block maximum surge levels.

In experiment E0, all GEV parameters are held at the baseline values and the 100-year storm surge is stationary. In experiment E1, we generate simulated annual block maxima nature runs using a nature state with nonstationary $\sigma$ and a stationary $\mu$. For simplicity, we construct a linear time-dependent change in $\sigma$ ($\sigma(t)$ in equation 3), to create a linear increase in the 100-year surge, where:

$$\sigma(t) = \sigma_o + \beta t. \qquad (3)$$

Experiment E2 employs a nature state with a stationary $\sigma$, and a linear increase in $\mu$ ($\mu(t)$ in equation 4), to create a linear change to the 100-year storm, where:

$$\mu(t) = \mu_o + \alpha t. \qquad (4)$$

The nature state for experiment E3, uses a nonstationary $\sigma$ and $\mu$, such that each parameter contributes half of the total increase in the 100-year storm surge rates (equation 5) to create a linear change to the 100-year storm, where:

$$\sigma(t) = \sigma_o + \beta t; \; \mu(t) = \mu_o + \alpha t. \qquad (5)$$

In experiment E4 we increase the 100-year surge by alternatively increasing $\mu$, and $\sigma$ in an 80 year cycle. For the first 40 years of the cycle we increase $\mu$, holding $\sigma$ constant, then, for the next 40 years, we hold $\mu$ constant while increasing $\sigma$.

In all experiments, we use four analysis methods (A-D) corresponding to each version of the GEV distribution to generate simulated observations of estimated 100-year surge levels (Gilleland and Katz 2016). Method A assumes fully stationary parameters. Method B assumes a nonstationary $\sigma$, with stationary $\mu$. Method C assumes a stationary $\sigma$, and nonstationary $\mu$. Method D assumes $\sigma$, and $\mu$ are nonstationary. For every nature run of every experiment, we use all four methods to estimate the effective 100-year return level for each multi-decadal interval (0-10, 0-20, 0-30 year et cetera). That is, we increase the length of the time series by adding an additional decade of nature run annual block maxima.

We use estimated 100-year storm surges from E0 to establish detection thresholds corresponding to IPCC conventional usage of *likely*, *very likely*, and *extremely likely* to characterize the likelihood of detecting a change in the estimated 100-year surge in simulated observations. For each detection analysis method (A-D), we calculate the 66%, 90%, and 95% quantile levels of the estimated 100-year surge across every simulation year, resulting in twelve threshold curves. Using the 95% threshold, for example, we can assign a 5% probability that a calculated 100-year surge is above the threshold by chance, thus we are 95% confident that the observation reflects an actual increase in the 100-year storm surge. From each experiment's nature runs we use detection methods A-D to generate simulated observations for each multi-decadal interval. We declare a successful detection of an increased 100-year surge when the estimated 100-year surge for a multi-decadal interval calculated for each analysis method exceeds the threshold for the corresponding analysis method (see Fig. 2). We calculate the percentage of successful detections that occur for every case of each experiment, (E1-E4), using each detection analysis method, (A-D), at each multi-decadal interval and for each likelihood level. Here we describe results from the one meter per century case examined at the 95% (extremely likely) detection thresholds. Additional cases are shown in the supplemental.

## 3. Results

We find, as expected, that longer observation periods typically increase the reliability of detection and reduce the frequency of false detections. Detection times are usually shorter and detections are more reliable when using nonstationary analysis methods (with key caveats discussed below). Estimated 100-year surge height simulated observations do not necessarily increase monotonically (Fig. 2). In many cases early detections are reversed in subsequent decades. In the one meter per century cases (Fig. 3), earliest median detections occur between 90 and 120 years. The bias and uncertainty associated with the 100-year storm surge estimated at the 100-year time frame is different for each combination of experiments (E1-E4) and detection method (A-D) (Fig. 4).

In experiment E0, where we test our observation systems against a static nature state with a constant 100-year surge and fully stationary GEV parameters, all detection methods recover a detection rate equal to one minus the likelihood level, consistent with the null hypothesis of stationary storm surges.

We next consider experiment E1, where the generating function uses a nonstationary $\sigma$ and a stationary $\mu$ (Fig. 3a). Using the stationary analysis method (A) detection of change to the 100-year surge is quite slow. After 100 years of additional observations method A only detects (at the 95% likelihood level) that the 100-year surge has increased about 25% of the time. Adding a nonstationary $\mu$ (C) to our detection model results in minor improvements to detection times. As perhaps expected, switching to a nonstationary $\sigma$ with either a stationary (B) or a nonstationary (D) $\mu$ detection model substantially decreases detection times.

In experiment E2, with a stationary $\sigma$ and a nonstationary $\mu$ (Fig. 3b), method A produces a counterintuitive result. Detection frequencies are initially lower (less than 5% for the very likely scenario at 50 years) than detection frequencies associated with a static nature state (indicating fewer detections than that of our fully stationary case used for testing the null hypothesis). This occurs when the best fit of the GEV equation to the nature runs results in a lower $\sigma$ or $\xi$, which reduces the 100-year storm surge. This effect is even more pronounced for method B. At the likely level (see supplemental for additional figures and discussion) and after 50 years, detection of the increasing 100-year surge occurs about 20% of the time using method B, but false detection of a decrease in the 100-year surge occurs approximately 40% of the time, an example of negative learning (O'Neill et al. 2006; Oppenheimer et al. 2008). The earliest and most reliable detections are achieved when using a stationary $\sigma$ and nonstationary $\mu$ (C). Using a fully nonstationary analysis (D) slightly degrades the reliability of detection.

In experiment E3 and E4, method D provides the best approach to detect changes in the 100-year events in these particular nature states, where the linear increase in 100-year surge is driven by linear or nonlinear changes to $\sigma$ and $\mu$ (Figs 3c and 3d). In all experiments, we calculate how frequently we predict a statistically significant increase in the risk of the 100-year surge and are able to identify the appropriate method for detection based on the original generating function.

The supplement contains additional figures for 66%, 90%, and 95% likelihood and 0.5 1.5, and 2.0 m/century rates. We have not considered the uncertainty of our initial 100-year surge estimate.

Long detection times may pose problems for decisionmakers who would assume- the availability of clear signposts to inform adaptive risk management strategies. As an alternative to analyzing detection times, we can consider how estimated 100-year surge levels evolve over time (Fig. 4). A desired characteristic of detection methods is that additional observations will cause estimates to converge towards the actual 100-year surge level. With method A, this does not occur. Longer observation timeframes increase the tendency to underestimate the 100-year storm surge. By using a fully stationary analysis for a nonstationary case, we identify an increasing bias with more observations for all our experiments. Nonstationary methods B-D converge towards the 100-year storm surge for experiments E1-E3, though at different rates. Additionally, the distribution of estimated storm surges differs for each combination of experiment and method. In experiment E4 (generated from nonlinear changes to $\sigma$ and $\mu$), none of the detection methods converge to the actual 100-year surge over the 200-year simulated timespans. This evolution in bias is discussed in the supplement and illustrated in Fig. S4.

## 4. Discussion

The risks of extreme events in the US are often summarized using a single 100-year return level (FEMA 2005, 2015). This metric is often communicated, by drawing a line on a map, without explicit estimates of the uncertainties (FEMA 2005, 2015). Other studies, while not directly addressing potential changes to the 100-year storm surge, suggest that the large increases in the 100-year surges of the magnitudes considered in this study may be possible at The Battery (Coch 1994; Scileppi and Donnelly 2007; Emanuel and Ravela 2013; Talke et al. 2014b; Reed et al. 2015). When realized through a change to the location, scale, or shape parameter (or combinations of the three), changes to the 100-year surge can drive considerable changes in flooding risks.

Our analysis shows that for the considered OSSE nature states, when changes in the 100-year surge are caused by a changing $\mu$, detection occurs earlier (on average) compared to the same shift in the 100-year surge caused by a changing σ. Moreover, the detectability of changes to the 100-year surge depends on the detection analysis method used, where the best results are obtained when the detection method matches what is actually occurring (i.e. as used in the nature state parameter change mechanisms). In considering experiment E2-E4 results we find that differences between the nature state and detection method can result in fewer detections than what is expected when the 100-year surge is not increasing (our null hypothesis cases). Additionally, we find that the distribution of estimated 100-year surge heights resulting from a prescribed change in the nature state depends upon both the analysis method and how the nature state statistical parameters are changing.

In our OSSE experiments E1-E3, we only consider nature states with linear changes to the 100-year surge driven by linear changes to $\sigma$ or $\mu$, excluding the possibility of a nonstationary $\xi$, and, in E4, consider one simple set of nonlinear changes to $\sigma$ and $\mu$. In the real world, actual physical phenomena may drive any combination of nonstationary GEV parameter forcings, introducing further sources of uncertainty into both the time required for detection of any change to the 100-year surge and our future estimates of the 100-year surge.

In stepping from our OSSE to the real world, expected changes to extreme weather events, including tropical and extratropical cyclones, could lead to alterations in storm frequency, intensity, size or track (Emanuel 1987; Beniston et al. 2007; Knutson et al. 2008; Bender et al. 2010; Reed et al. 2015; Little et al. 2015; Kossin 2017). These may, in turn, drive nonlinear changes to any of the GEV parameters. As discussed above, a mismatch between the underlying phenomena and the analysis method can impact the accuracy of analysis or result in negative learning. In analyzing real world observations, analysts rarely have the luxury of knowing how reality, in terms of these underlying parameters, is changing. Therefore, it is easy to imagine scenarios where long-term nonlinear trends in storm characteristics drive such mismatches between reality and the chosen statistical model for the detection. This can lead to overconfident or biased projections of future storm surge risk that can result in poor decisions.

Various strategies have been proposed to select and manage policies and infrastructure investments required to manage storm surge risk (e.g., van Dantzig 1956; Kok and Hoekstra 2008; Linquiti and Vonortas 2012). In addition, it has long been known that storm surge exceedance probabilities and associated risk may not be constant over time (e.g., van Dantzig 1956, Lempert et al, 2012). Hence, it is important to learn from experience and adapt flood prevention strategies. An often used strategy is to set dike heights to the return level for a prescribed annual exceedance probability (e.g., one in 100 years in the US, or longer return periods in the Netherlands) as estimated at the time of design with an additional fixed safety margin (e.g., FEMA 2005; Kok and Hoekstra 2008; Ligtvoet et al. 2012; Zevenbergen et al. 2013). Older strategies rely on resetting maximum dike heights based on new maximum observations at some time interval (Kok and Hoekstra 2008). These strategies can result in similar overtopping and overall economic performance (Kok and Hoekstra 2008; Linquiti and Vonortas 2012), but, strategy performance can depend on the ability to learn from observations and to adapt strategies over time (Linquiti and Vonortas 2012). The longer it takes to learn about increasing risk, the more delayed adaptation responses may be, which could result in unanticipated risk increases.

When changes to risk mitigation strategies (such as NYC proposals to implement adaptive risk management strategies) are considered, the estimated 100-year storm surge is often used as a line-in-the-sand demarcation of risk, sometimes with a freeboard allowance to incorporate additional uncertainty (FEMA 2005, 2015; NYC Economic Development Corporation 2014; Kaplan et al. 2016). For a given set of GEV parameters, the statistical uncertainty surrounding return level estimates can also be estimated and can be quite large (e.g.,

Coles 2001). This estimate of uncertainty, however, often does not account for the increased uncertainty described here, introduced by both statistical mechanisms driving changes to the 100-year storm surge and mismatches between the detection model and reality. Stated another way, uncertainty resulting from mismatches between reality and our detection model may underestimate the uncertainty in both the actual level of our 100-year storm surge estimates and the time required to notice a change to that level.

Developing public consensus on the need for reducing storm surge risk can be a slow process. Even after consensus is reached, evaluating options, designing strategies, establishing public financing, planning, and implementing major construction projects adds delays between identification of the need for defenses and their implementation. Adopting an adaptive risk management or dynamic adaptive policy pathway can substantially reduce initial investment cost, potentially reducing the lag time associated with subsequently implementing defensive strategies within a previously agreed framework (Lempert et al. 1996; Kwakkel et al. 2015). The success of these strategies, however, depends on the detection of signposts signaling the need to adapt or switch policies (Kwakkel et al. 2015). When considering storm surge risk management strategies, failing to account for long detection time frames and the potential for contra-indicating results and negative learning can degrade the performance of adaptive strategies.

At the start of this paper, we posed the simple question: How fast can we learn from past and potential future storm surge observations about potential changes? The answer is simple. It can take a long time. Quantitatively, however, the answer is more complicated. We show the answer (in Figs. 2-4) for a very limited set of changing parameters, at a few likelihood levels and for a nature state where we prescribe statistical parameters that we selected for generating the changes. The additional uncertainty associated with any particular combination of these circumstances may not be captured in traditional extreme value analyses, thus potentially leading to underestimated uncertainty. The additional uncertainty for any single set of changing parameters can be probabilistically estimated using the methods outlined in this paper and detailed in the supplementary materials. In the real world, analysts and decisionmakers often do not have this luxury, and the problem can be more difficult. Nevertheless, decision analyses may be biased if they neglect the possibility of overly confident estimates of future risk that could occur if the additional structural uncertainties outlined here are not considered. Failing to do so may lead to poor adaptive management strategies and, when new protective strategies are implemented, a failure to meet desired protection levels.

**Acknowledgements**

We thank M. Tingley, M. Haran, B. Lee, G. Garner, and R. Lempert for useful discussions. This research was partially supported by the National Science Foundation (NSF) through the Network for Sustainable Climate Risk Management (SCRiM) under NSF cooperative agreement GEO-1240507 and the Penn State Center for Climate Risk Management. Any opinions, findings, and conclusions or recommendations expressed in this material are those of the authors and do not necessarily reflect the views of the NSF.

**Code and data availability**

The data and code are currently available upon request from the corresponding author.

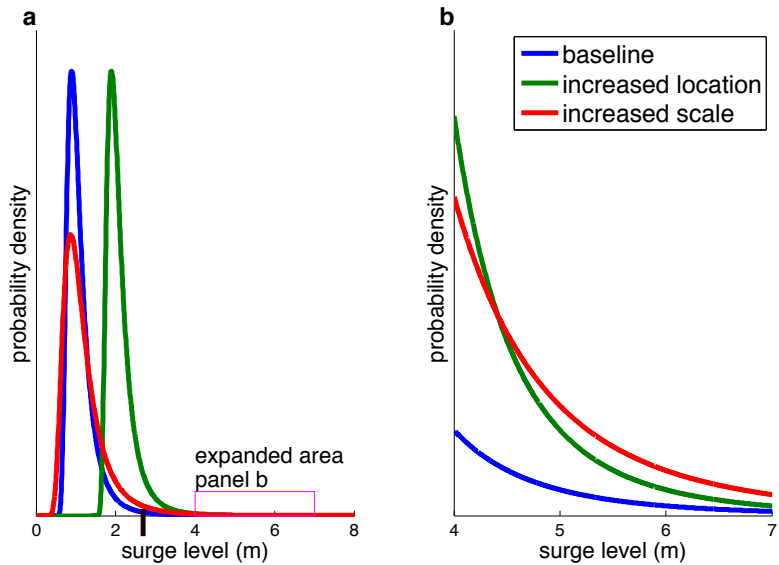

Figure 1. The GEV distribution estimated using the historic data observed at *The Battery* from 1924 to 2014 (blue solid lines). The left panel shows the probability density (vertical axis) of highest annual block maximum surges from 0-8 meters (horizontal axis). The right plot is an expanded view on the left panel magenta region, showing the surge range between 4 to 7 meters. The green and red lines show the change to the pdf that results from changing the scale and location parameters respectively to achieve a one meter increase in the hundred-year storm surge. The large black tick at 2.6 meters marks the height of the highest surge recorded at *The Battery* during super-storm Sandy.

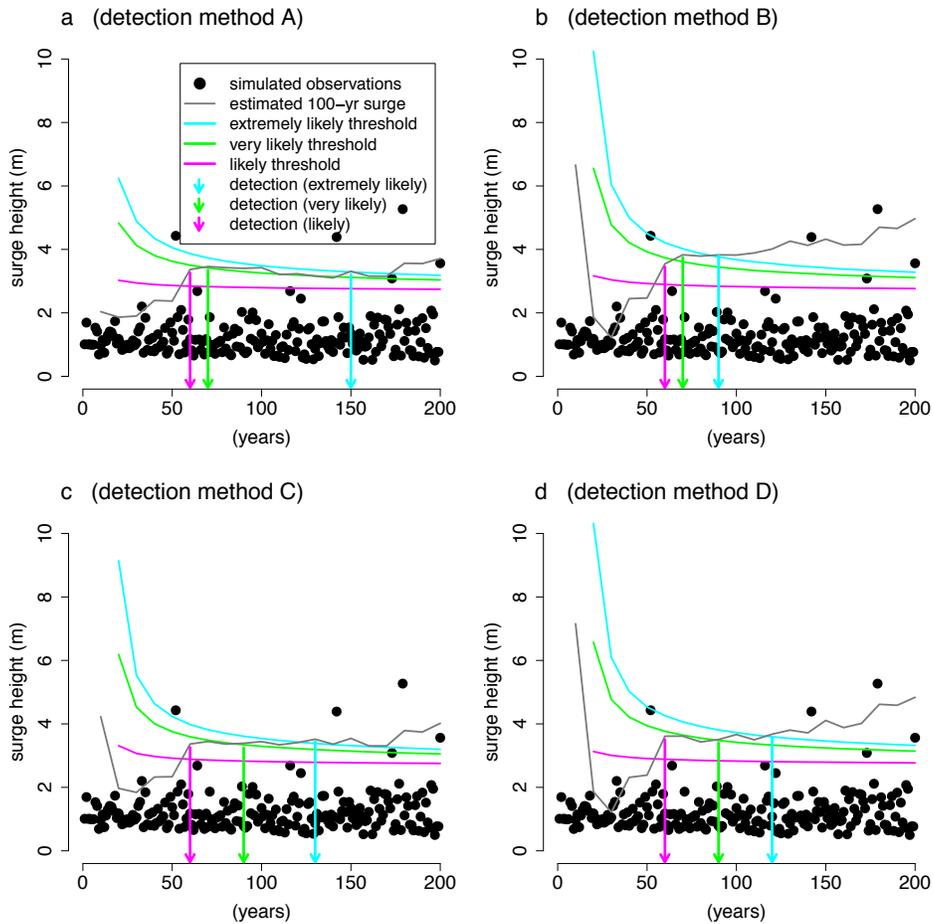

Figure 2. Threshold estimates for 100-year storm surge events with detection examples. All four plots are based on one 200-year nature run from experiment E1 (increasing scale parameter) case 1 (1m/century increase in the 100-year storm surge). Nature run annual block maximums (ABM) are shown by black dots. Each panel (a-d) displays thresholds, the estimated 100-year storm surge simulated observations, and detection times for each respective detection method (a, fully stationary, b, non-stationary scale, stationary location, c, stationary scale, non-stationary location, and non-stationary Scale, non-stationary location. Detection thresholds for the respective detection methods are shown for extremely likely (cyan), very likely (green), and likely (magenta).) Detection of a change to the 100-year storm surge at a given confidence occurs when the estimated 100-year storm surge simulated observation (calculated at each 0-10, 0-20, etc. decadal interval) exceeds the associated threshold. The decade of initial detection is indicated by arrows whose color corresponds to the associated confidence level.

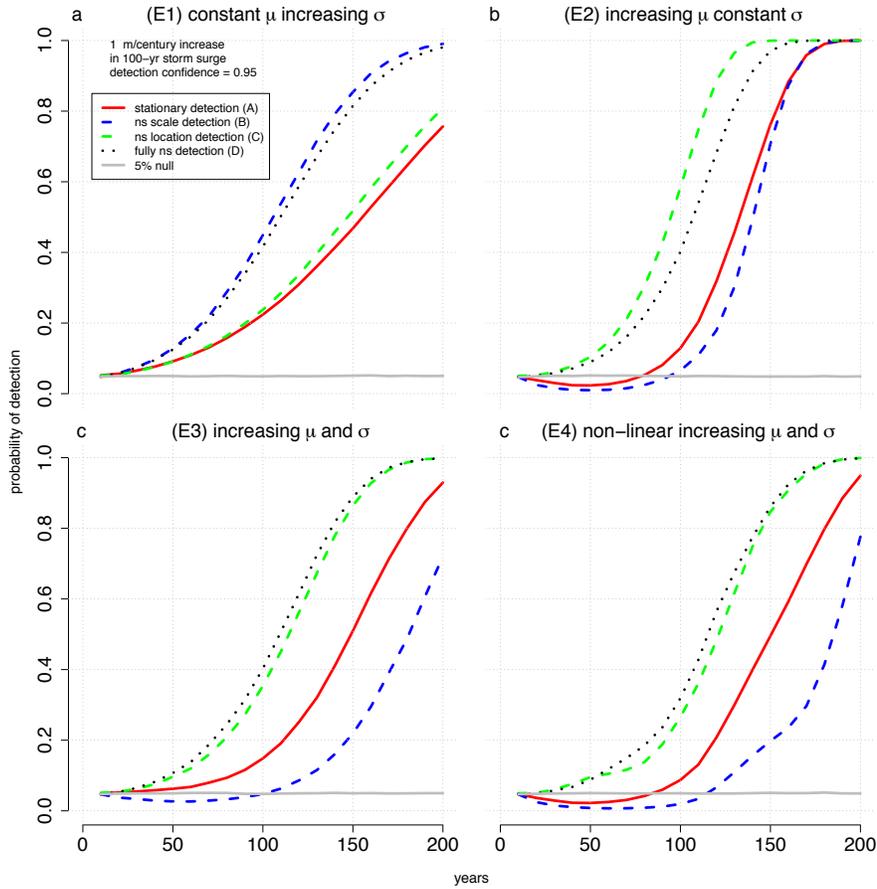

Figure 3. Frequency of detection results (one m increase per century case) for the four experiments comparing alternative detection analysis methods. The non-stationary scale parameter (E1), non-stationary location parameter (E2), non-stationary scale and location parameter (E3), and non-linear non-stationary scale and location parameter (E4) experiment results are shown in panels a-d respectively. The percentage of successful detections achieved for fully stationary analysis (method A, solid red), non-stationary scale parameter, $\sigma$, analysis (method B, dashed blue), non-stationary location, $\mu$, analysis (method C, dashed green), and non-stationary $\sigma$ and $\mu$ analysis (method D, dotted black), is plotted for each multi-decadal interval. The gray line shows the frequency of detection rate for fully stationary parameters.

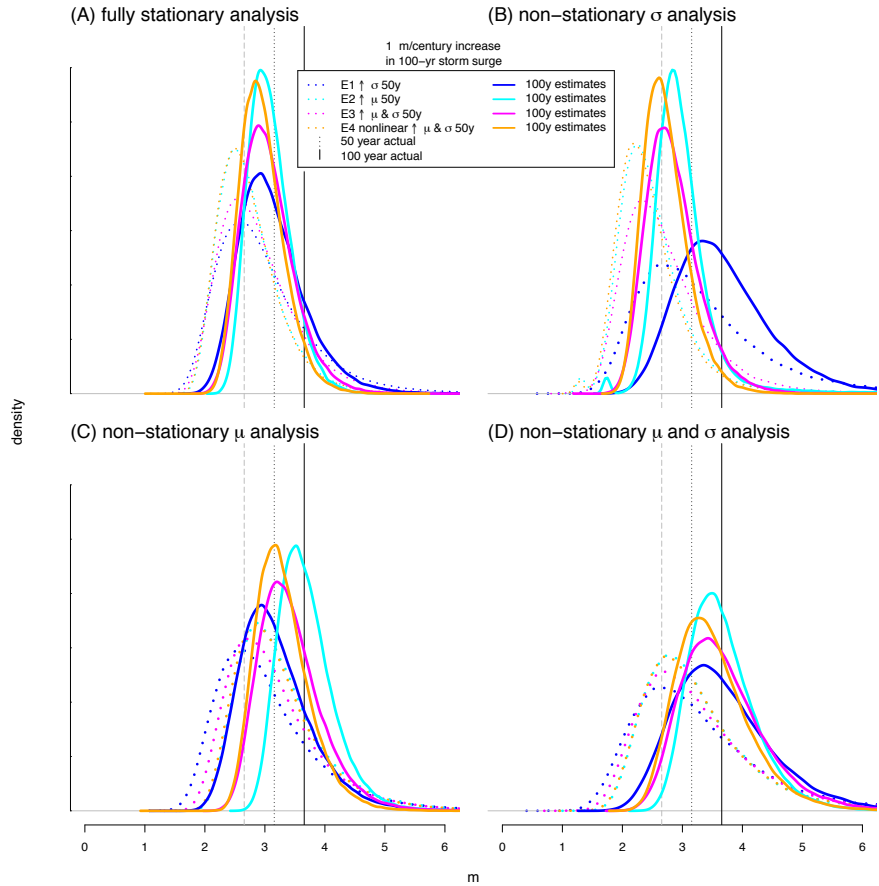

Figure 4. Probability density of simulated 100-year surge observations for each detection method (A-D) for each experiment (E1-E4). The stationary (A), non-stationary $\sigma$, (B), non-stationary location, $\mu$ (C), and non-stationary $\sigma$ and $\mu$ (D) detection methods are displayed in panels A-D respectively. For each method, the probability density of 100 year surge ssimulated observations for non-stationary $\sigma$, (E1, blue ), non-stationary location, $\mu$ (E2, cyan), non-stationary $\sigma$ and $\mu$ (E3, magenta), and non-linear non-stationary $\sigma$ and $\mu$ (E4, orange) is plotted for 0-50 and 0-100 decadal intervals. Initial, 50 year, and 100 year nature state 100-year surge levels indicated by grey dashed, dark grey dotted, and solid black vertical lines respectively.